\documentclass[12pt]{article}
\usepackage[dvips]{graphicx}
\begin{document}
\title{Decay of charged scalar field around a black hole: quasinormal modes
of R-N, R-N-AdS and dilaton black holes}
\author{R.A.Konoplya \\
Department of Physics, Dniepropetrovsk National University\\
St. Naukova 13, Dniepropetrovsk  49050, Ukraine\\
konoplya@ff.dsu.dp.ua}
\date{}
\maketitle
\thispagestyle{empty}
\begin{abstract}
It is well known that the charged scalar perturbations of the
Reissner-Nordstrom metric will decay slower at very late times
than the neutral ones, thereby dominating in the late time signal.
We show that at the stage of quasinormal ringing, on the contrary,
the neutral perturbations will decay slower for RN, RNAdS and dilaton
black holes. The QN frequencies of the nearly extreme RN
black hole
have the same imaginary parts (damping times) for charged and neutral
perturbations. An explanation of this fact is not clear
but, possibly, is connected with the Choptuik scaling.

\end{abstract}

\section{Introduction}

The response of a Schwarzschild black hole as a Gaussian wave packet
impinges upon it consists of decaying quasinormal oscillations, dominating
after time $t \approx 70 M$, and inverse power-low tails, dominating
after time $t \approx 300 M$, where $M$ is the black hole mass
(see \cite{Kokkotas-Schmidt} and references therein). The quasinormal
ringing can be caused by either external fields or by the formation
of a black hole itself, and, the characteristic frequencies do not
depend on a form of perturbations, giving us a "footprint" of a black
hole.

Due to AdS/CFT correspondence \cite{Maldacena} the investigation of the quasinormal
frequencies of AdS black holes is appealing now: it gives the
thermalization time scale for a field perturbation \cite{Horowitz1},
namely, the imaginary part of the quasinormal frequency, being inversely
proportional to the damping time of a given mode, determines the
relaxation time of a field. Thus the more imaginary part of
$\omega$ the faster a given field comes to an equilibrium.

The investigation of the QN modes within the ADS/CFT correspondence
was initiated on the ADS gravity side by Horowitz and Habeny
in \cite{Horowitz1} for
massless scalar field.
Then quasinormal modes associated with perturbations of
different fields were considered in a lot of works \cite{Cardoso-Lemos1}-
\cite{Moss-Norman}.
An exact expression for the three-dimensional BTZ black hole QN modes
corresponding to fields of different spin was obtained
by Cardoso and Lemos in \cite{Cardoso-Lemos1}.
Recently the similar work for the BTZ black hole was done on the CFT side
\cite{Birmingham2}.

On the AdS gravity side, it was found that for
the neutral massless scalar field in the background of
RNAdS black hole with small charge, the more the black hole charge
is,
the quicker for its approach to thermal equilibrium in CFT
\cite{Wang2}, and after the black hole charge approaches some critical
value the situation changes on contrary \cite{Wang1}.
This repeats the behaviour of the usual R-N quasinormal spectrum, where
the imaginary part of $\omega$ grows with the black hole charge
up to some maximum, and then begin to decrease.

Summarizing the results of the papers \cite{Wang1},
\cite{Brady-Chambers1}, \cite{Brady-Chambers2}, and \cite{Price}, one
can see that the late time radiative behaviour of a neutral scalar
field for asymptotically flat (R-N) and asymptotically
(anti)-de-Sitter (RNdS and RNAdS) black hole space-times is
essentially different: in the first case the inverse power-law tails
are dominating, while in the second it is an exponential
decay. This decay is oscillatory for RNAdS, and for RNdS when
a scalar field strongly coupled to curvature.

When collapsing a charged matter, a charged black hole forms.
Thus the evolution of a charged scalar field outside the R-N black
hole is a most relevant. The late time behaviour of a charged scalar
field was considered by S.Hod and T.Piran  \cite{Hod-Piran1}.
There was shown that in the radiative tails the neutral
perturbations decay faster than the
charged ones and therefore dominate at very late times. In addition,
while at timelike and null infinity inverse power-law tails appear,
along the future black hole event horizon, an oscillatory behaviour
accompanies this tail.

Thus there is a quite clear picture of asymptotic behaviour of
the radiation corresponding to charged perturbations,
while its behaviour during
the stage of quasinormal ringing is lacking.
This motivated us to study the behaviour of a complex (charged)
scalar field during the quasinormal ringing through the computing of its
resonant characteristic frequencies for R-N and R-N AdS black holes.
In Sec.2 we shall compute the
quasinormal frequencies of the R-N black hole for
different multipole numbers $l$, in Sec.3 the case of
the R-N-AdS  black hole is  considered, and in Sec.4
the dilaton black QN frequencies are obtained.
We have found that the modes of the nearly extremal R-N
black holes have
the same damping times for charged and neutral
perturbations .
The
possible connection of this fact with the critical collapse is
discussed in Sec.4.

\section{Reissner-Nordsrom black hole}

We shall consider the evolution of the charged scalar perturbations
field in the background of the Reissner-Nordstrom metric:
\begin{equation}\label{1}
ds^{2}= -f(r) dt^{2} + f^{-1}(r)dr^{2} +r^{2}d\Omega^{2}_{2},
\end{equation}
where $ f(r)= 1-\frac{2 M}{r}+\frac{Q^2}{r^2}$. The wave equation of
the complex scalar field has the form:
\begin{equation}\label{2}
\phi_{;ab} g^{ab}-i e A_{a} g^{ab}(2 \phi_{; b}-i e A_{b}
\phi)-i e A_{a; b} g^{ab} \phi =0,
\end{equation}
here the electromagnetic potential $A_{t} = C -\frac{Q}{r}$, $C$ is a
constant. After representation of the charged scalar field into spherical
harmonics and some algebra the equation of motion takes the form
\cite{Hod-Piran1}:
\begin{equation}\label{3}
\psi_{,tt}+2i e \frac{Q}{r} \psi_{,t} - \psi_{, r^{*}r^{*}}+ V
\psi=0,
\end{equation}
where
\begin{equation}\label{4}
V = f(r)  \left(\frac{l(l+1)}{r^2} +\frac{2 M}{r^3} -
\frac{2 Q^2}{r^4}\right) -e^{2} \frac{Q^2}{r^2},
\end{equation}
and $\psi = \psi (r) e^{-i \omega t}$.
One can compute the quasinormal frequencies stipulated by the above
potential by using the third order WKB formula of S.Iyer and C.Will
\cite{Will1}:
\begin{equation}\label{5}
\frac{i Q_{0}}{\sqrt{2 Q_{0}''}}
-\Lambda(n)-\Omega(n)=n+\frac{1}{2},
\end{equation}
where $\Lambda(n)$, $\Omega(n)$ are second and third order WKB
correction terms depending on the potential $V$ and its derivatives
in the maximum. Here $Q = -V+\omega^{2} - 2 \frac{e Q}{r} \omega$.
Since $Q$ depends on $\omega$, the procedure of finding of the QN
frequencies is the following: one fixes all the parameter of the
QN frequency, namely, the multipole index $l$, the overtone number
$n$, the black hole mass and charge $M$ and $Q$, and $e$; then
one finds the value of $r$ at which $V$ attains a maximum as a
numerical function of $\omega$ and substituting it into the formula
(\ref{5}) one finds, with the trial and error way, $\omega$ which
satisfies the equation (\ref{5}).
\begin{table}
\begin{center}
\begin{tabular}{ccccc}
\hline
  $Q$ & $l=1$ & $l=2$ \\
\hline
  0   & $0.2911 - 0.0980 i$ & $0.4832 - 0.0968 i$ \\
      & $0.2911 - 0.0980 i$ & $0.4832 - 0.0968 i$ \\
\hline
  0.1 & $0.2916 - 0.0981 i$ & $0.4840 - 0.0969 i$ \\
      & $0.2951 - 0.0984 i$ & $0.4874 - 0.0971 i$ \\
\hline
  0.3 & $0.2958 - 0.0984 i$ & $0.4908 - 0.0973 i$ \\
      & $0.3064 - 0.0995 i$ & $0.5011 - 0.0979 i$ \\
\hline
  0.5 & $0.3049 - 0.0991 i$ & $0.5056 - 0.0980 i$ \\
      & $0.3233 - 0.1008 i$ & $0.5236 - 0.0990 i$ \\
\hline
  0.7 & $0.3212 - 0.0996 i$ & $0.5322 - 0.0986 i$ \\
      & $0.3490 - 0.1018 i$ & $0.5593 - 0.1000 i$ \\
\hline
  0.8 & $0.3337 - 0.0992 i$ & $0.5527 - 0.0983 i$ \\
      & $0.3672 - 0.1015 i$ & $0.5855 - 0.0999 i$ \\
\hline
  0.9 & $0.3509 - 0.0972 i$ & $0.5815 - 0.0966 i$ \\
      & $0.3918 - 0.0993 i$ & $0.6214 - 0.0980 i$ \\
\hline
  0.95 & $0.3622 - 0.0946 i$ & $0.6011 - 0.0945 i$ \\
       & $0.4079 - 0.0960 i$ & $0.6459 - 0.0952 i$ \\
\hline
  0.99 & $0.3729 - 0.0907 i$ & $0.6205 - 0.0902 i$ \\
       & $0.4231 - 0.0908 i$ & $0.6701 - 0.0904 i$ \\ \hline
\end{tabular}
\end{center}
\caption{The quasinormal frequencies for RN BH, $l=1,2$ $n=0$, $e=0$
(first line)
and $e=0.1$ (second line).}
\end{table}

It is essential that for  $l=0$ modes the WKB formula
gives the worse precision: a relative error, for example,
for scalar perturbations of Schwarzschild BH, may be of order
$10$ per cents \cite{Will2}.
Nevertheless the more $l$ (and the less $n$), the more accurate WKB formula
is, and already for $l=3$, $n=0$, according to general experience,
a relative error may be of order $10^{-2}$ per cents. Thus in order to be
sure that not only the WKB frequencies of charged and neutral
perturbations coincide in the extremal limit,
but also the true frequencies do the same, one needs to proceed
computations to higher $l$.
We can see it from the Fig.2-3, where $l=3$ frequencies are presented.
For higher multipole indexes, precision is better.

\begin{figure}
\begin{center}
\includegraphics{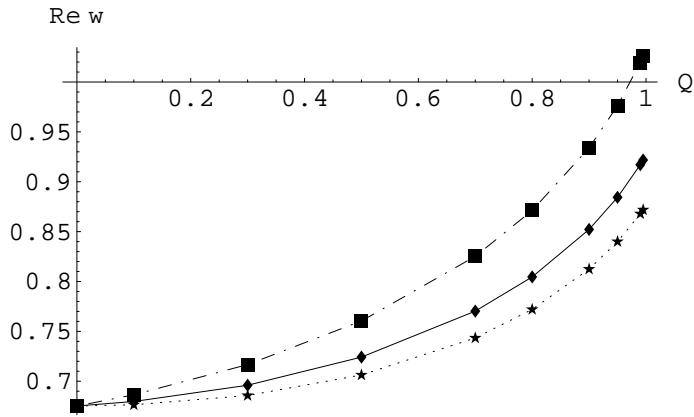}
\caption{Real part of $\omega$, $l=3$, $n=0$, $e=0$ (star), $e=0.1$
(diamond) and
$e=0.3$ (box), for $Q$, running from $0$ to $0.995$ (R-N BH).}
\label{prd2_fig3}
\end{center}
\end{figure}

\begin{figure}
\begin{center}
\includegraphics{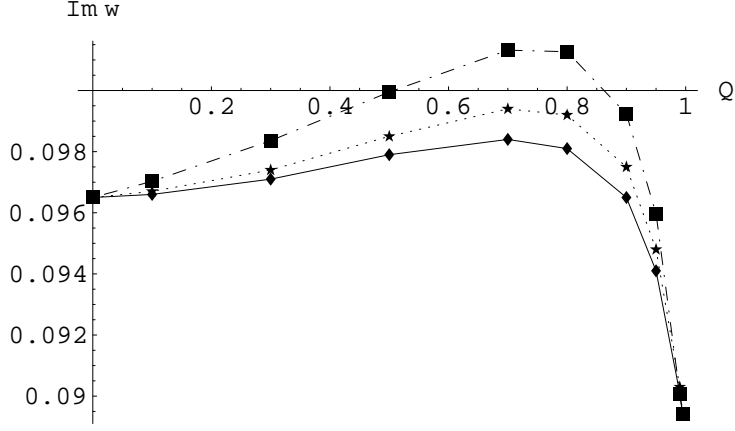}
\caption{Imaginary part of $\omega$, $l=3$, $n=0$, $e=0$ (diamond), $e=0.1$
(star) and
$e=0.3$ (box),  for $Q$, running from $0$ to $0.995$ (R-N BH).
At $Q=0.995$ $\omega_{Im}=0.08938$ for $e=0$ and $\omega_{Im}=0.08943$
for $e=0.3$, which is within precision of the WKB method.}
\label{prd2_fig4}
\end{center}
\end{figure}

Real part of $\omega$ for both neutral and charged scalar fields
grows with increasing of charge $Q$, $\omega_{Im}$ is more for
charged perturbations than for a neutral one.
In addition, and this is the most interesting feature of charged
QN spectrum, the imaginary part of a given "charged mode"
approaches the neutral one in the limit of the extremal black hole.
Within the third order WKB method one can check it with a high accuracy
for higher multipole number perturbations.

\section{Reissner-Nordsrom-Anti-de-Sitter black hole}

The Reissner-Anti-de-Sitter metric has the form
\begin{equation}\label{100}
ds^{2}= -f(r) dt^{2} + f^{-1}(r)dr^{2} +r^{2}d\Omega^{2},
\end{equation}
where
\begin{equation}\label{101}
f(r)=1-\frac{r_{+}}{r}
-\frac{r^{3}_{+}}{r R^{2}}- \frac{Q^2}{r r_{+}} +\frac{Q^2}{r^2}+
\frac{r^2}{R^2}.
\end{equation}

Quasinormal oscillations associated with the decay of the charged
scalar field in the background of RNAdS are governed by the wave equation
(\ref{3}), which can be transformed to the form
\begin{equation}\label{6}
f(r) \frac{d^{2}\psi(r)}{d r^{2}} + (f'(r) -2 i \omega)
 \frac{d \psi(r)}{d r} - V(r) \psi (r)=0.
\end{equation}
Here the $V(r)$ is, again, a frequency dependent potential, determined
by the formula:
\begin{equation}\label{7}
V(r)= \frac{f'(r)}{r} +\frac{l (l+1)}{r^2} +
\frac{1}{f(r)} \left(2 \frac{e Q}{r} \omega -e^{2} \frac{Q^2}{r^2}\right)
\end{equation}
By rescaling of $r$ we can put $R=1$. The effective
potential is infinite at spatial infinity. Thus the wave function is
considered to vanish at infinity and satisfies the purely in-going
wave condition at the black hole horizon.
Then one can compute the quasinormal frequencies stipulated by the
potential (\ref{7}) following the procedure of G.Horowitz and V.Habeny
\cite{Horowitz1}. The main point of that approach is to
expand the solution to the wave equation (\ref{6}) around
$x_{+}=\frac{1}{r_{+}}$ ($x=1/r$):
\begin{equation}\label{8}
\psi (x)=\sum_{n=0}^{\infty} a_{n}(\omega)(x-x_{+})^{n}
\end{equation}
and to find the roots of the equation $\psi(x=0)=0$ following from
the boundary condition at infinity. In fact, one has to truncate the
sum (\ref{8}) at some large $n=N$ and check that for greater $n$ the
roots converge.

While the quasinormal modes of an asymptotically flat black hole are
proportional to its mass, those of an asymptotically anti-de-Sitter
black hole depend upon the radius of a black hole. For large
($r_{+}$ is much greater than the anti-de-Sitter radius $R$) and
intermediate Schwarzschild-Anti-de-Sitter black holes,
both $\omega_{Re}$ and  $\omega_{Im}$ are proportional to the black
hole temperature. For small black holes $r_{+} \ll R$ this linearity
breaks and in the limit $r_{+} \rightarrow 0$ the QNM approaches the pure
Anti-de-Sitter modes \cite{Horowitz1}, \cite{Konoplya1}.
Since it is a large black hole which is of direct interest for AdS/CFT
correspondence, we shall restrict ourselves to this black hole regime.

From the Fig.5,6 one can see that $\omega_{Re}$ and $\omega_{Im}$
grow with increasing of the charge conjugation $e$, i.e. the real
oscillation frequency is more for charged perturbations than for a
neutral ones and the damping time of a given mode is more for
neutral perturbations. Yet we managed to compute only the lowly
charged case, due to the two difficulties.
First, when $Q$ and $e$ grow, the number of terms in the
truncated sum representing the wave function $\psi$
increases: one has to sum over $N \sim 10^3$ and more,
and thus one has to guess new modes through  the trial and error way.
At the same time, the minimums of the truncated sum
corresponding to the
quasinormal frequencies are a most narrow in the $\omega$ plane and
one has to guess a lot of figures in the quasinormal frequency in
order to catch the above minimum.
Thus for highly charged
case one has to resort to another method of calculations of QN modes.

\begin{figure}
\begin{center}
\includegraphics{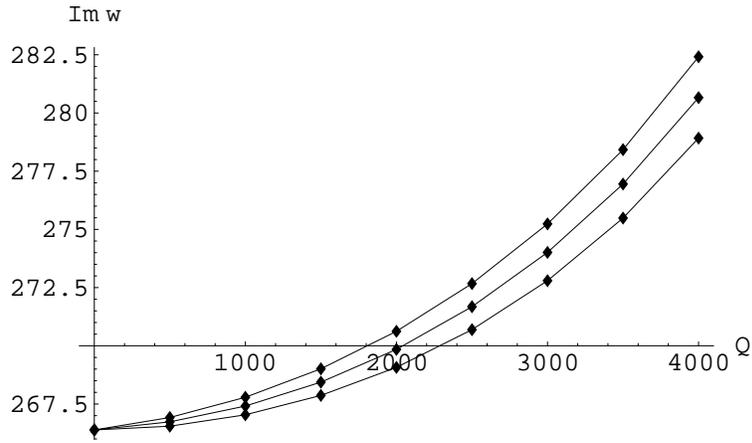}
\caption{Imaginary part of $\omega$, $l=0$, $n=0$ for $e=0$,  $e=5\cdot
10^{-5}$, and
$e=10^{-4}$ from the bottom to the top (R-NAdS BH).}
\label{prd2_fig5}
\end{center}
\end{figure}

\begin{figure}
\begin{center}
\includegraphics{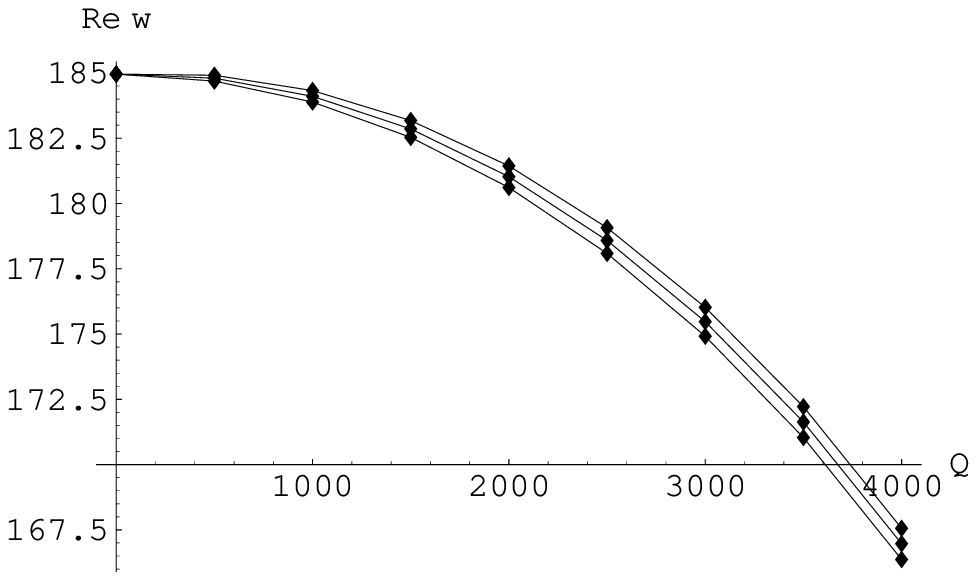}
\caption{Real part of $\omega$, $l=0$, $n=0$ for $e=0$,  $e=5\cdot10^{-5}$,
and $e=10^{-4}$ from the bottom to the top (R-NAdS BH).}
\label{prd2_fig6}
\end{center}
\end{figure}

\section{Dilaton black hole}

A wide class of theories includes the stationary spherically symmetric
black hole solution with massless scalar field of some specific form
(the so called dilaton):
\begin{equation}\label{9}
ds^{2} = \lambda^{2} dt^{2} - \lambda^{-2}dr^{2} -R^{2} d\theta^{2}
-R^{2} \sin^{2}\theta d\varphi^{2}
\end{equation}
where
\begin{equation}\label{10}
\lambda^{2}=\left(1-\frac{r_{+}}{r}\right)
\left(1-\frac{r_{-}}{r}\right)^{\frac{1-a^{2}}{1+a^{2}}},\qquad
R^{2}=r^{2}\left(1-\frac{r_{-}}{r}\right)^{\frac{2a^{2}}{1+a^{2}}},
\end{equation}
and
\begin{equation}\label{11}
2 M = r_{+} +\left(\frac{1-a^{2}}{1+a^{2}}\right) r_{-},\qquad
Q^{2} = \frac{r_{-}r_{+}}{1+a^{2}}.
\end{equation}
Here the dilaton and electromagnetic fields are given by the
formulas:
\begin{equation}\label{12}
e^{2 a \Phi}
=\left(1-\frac{r_{-}}{r}\right)^{\frac{2a^{2}}{1+a^{2}}}, \qquad
F_{tr} = \frac{e^{2 a \Phi} Q}{R^{2}},
\end{equation}
where $a$ is a non-negative dimensionless value representing coupling.
The case $a =0$ corresponds to the classical
Reissner-Nordstr\"om metric, the case $a=1$ is suggested by the low
energy limit of the superstring theory, and $a=\sqrt{3}$ corresponds to
the dimensionally reduced Kaluza-Klein black hole.

Following the above WKB method we shall compute here the quasinormal
modes corresponding to the neutral and charged massless scalar test
field. We do not consider interaction with the dilaton, i.e.
the scalar field simply propagates in the black hole background.

The wave function obeys the equation (\ref{3})
with the effective potential
\begin{equation}\label{14}
V(r)=\frac{R,_{r^{*} r^{*}}}{R} +\frac{l(l+1) \lambda^{2}}{R^2}-
e^{2} \frac{Q^2}{r^2}
\end{equation}
This potential is broadening near the extremal limit \cite{Holzey-Wilchek}.
\begin{table}
\begin{center}
\begin{tabular}{ccc}
\hline
$ Q$ & $e=0$ &  $e=0.1$ \\
\hline
$ 0$ & $0.67521- 0.09651 i$ &  $0.67521- 0.09651 i$ \\
$ 0.2$ & $0.67977-0.09673 i$ & $0.68650-0.09970 i$ \\
$ 0.4$ & $ 0.69411-0.09738 i$ &  $0.70768-0.09799 i$ \\
$ 0.6$ & $ 0.72045-0.09853 i$ & $ 0.74109-0.09941 i$ \\
$ 0.8$ & $0.76379-0.10028 i$ &  $0.79189-0.10138 i$ \\
$ 1.0$ & $0.83580-0.10273 i$ &  $0.87209-0.10398 i$\\
$ 1.1 $& $0.89116-0.10419 i$ & $ 0.93205-0.10549 i$ \\
$ 1.2 $& $0.97122-0.10559 i$ &  $1.01734-0.10691 i$ \\
$1.3 $& $1.10340-0.10582 i$ &  $1.15602-0.10711 i$ \\
$1.35 $& $1.21774-0.10357 i$ &  $1.27473-0.10478 i$\\
$1.4 $& $ 1.45670-0.08874 i$ &  $1.52076-0.08969 i$ \\ \hline
\end{tabular}
\end{center}
\caption{The quasinormal frequencies for $a=1$ dilaton black hole $l=3$,
$n=0$, $e=0$ and $e=0.1$.}
\end{table}

In case of dilaton BH both real and imaginary parts of $\omega$ grow with
increasing of either $Q$ or $e$.
Nevertheless the imaginary part of $\omega$
of the charged field does not approach that of the neutral one in the
nearly extremal regime. One can see it on example of $l=3$, $n=0$
modes where the WKB method gives reasonable accuracy.

Recently it has been obtained that at late times the neutral
scalar field falls off faster than the charged field in the dilaton BH
background \cite{Moderski-Rogatko}. Thus we see that for a dilaton
black hole the situation changes on contrary as well:
domination  of a neutral field during the quasinormal
ringing and of charged field at late
times.

\section{Discussion}

We have learnt here that the damping time of the quasinormal
oscillations associated with a charged scalar field in the background
of the Reissner-Nordstrom black hole is less than that of a neutral
one. Thus that is the neutral perturbations which will dominate at
later stages of quasinormal ringing. Yet, we know that at late times
the charged perturbations are dominating, and one could expect,
possibly, that, the same sort of perturbations must dominate
in the earlier stages of radiation. The logic of the process,
however, is different. As was shown in \cite{Hod-Piran1}, the late
time behaviour of the charged scalar field is entirely determined by
the flat space-time effects, while that of the neutral perturbations
is depend on the relation between the "tortoise" $r^{*}$ coordinate
and $r$, i.e. by the space-time curvature. In other words, the
radiative tail of the charged field arises due to the backscattering
of this  field off the electromagnetic potential far away from the
black hole, while in the case of the neutral fields  it is the
effects of gravitation near the black hole (curvature effects).
In this context it seems natural that in the earlier periods of
radiation (quasinormal ringing) the curvature effects are dominating,
and the neutral perturbations will damp slower.

Another interesting point of this study is the coincidence of the
imaginary parts of $\omega$ for charged and neutral perturbations
for the nearly extremal black hole.
It leaps to the eyes at once
that since the universal index appearing in the phenomena of
critical collapse
$\beta$ equals $0.37$ both for charged \cite{Hod-Piran2} and
neutral  \cite{Choptuik}  scalar fields, then there may be a connection
between the behavior of the quasinormal spectrum of nearly extremal
black holes and the critical exponent for a given black hole.
Yet the latter conjecture seems be too strong, and the black hole
quasinormal modes may be related to the critical exponent in some
specific space-time geometries \cite{Horowitz1},\cite{Kim}.

In this connection it is interesting to remind that the nearly
extremal R-N black hole is effectively
described by the $AdS_{2}$
black hole after spherically symmetric dimensional reduction
\cite{Strominger}. For
such a reduced nearly extremal black hole an exact relation between
the quasinormal
modes and the critical exponent is obtained in
\cite{Kim}.

Yet the coincidence of the damping times for charged and neutral
modes of the neraly extremal black hole is, apparently, an exclusive
property of RN BH, and is not appropriate to other black holes.
It is possible, also that this coincidence takes place only
for a massless scalar filed, since for a massive one the situation
is qualitatively different \cite{Wang3}

Thus any kind of satisfactory explanation of the above coincidence
from physical point of view is lacking.

Whether the
damping times of charged and neutral perturbations in the nearly
extremal limit will coincide for RNAdS
BH, and, for an asymptotically non-flat black hole in general, is a
question for further investigation.

\section{Acknowledgments}
I wish to thank S.Ulanov and A.Gulov for invaluable discussions,
and F.Piazza for pointing out Ref. \cite{Piazza}.

\end{document}